%% file: DFPP_LOPSTR-17.tex
\documentclass[english,a4paper]{llncs}
\usepackage[latin1]{inputenc}
\usepackage[T1]{fontenc}
\usepackage{babel}
\usepackage{graphicx} 
\usepackage{amsmath}
\usepackage{amsfonts}
\usepackage{amssymb}
\usepackage{multirow}
\usepackage{multicol}
\usepackage{url}
\usepackage{stmaryrd}
\usepackage{lmodern}  
\usepackage{xspace}
\usepackage{enumitem}
\usepackage{hyperref}
\usepackage{float}      
\usepackage{breakcites} 

\usepackage{proof}
\usepackage{makecell}

\usepackage{tikz}
\usetikzlibrary{positioning}
\usetikzlibrary{arrows}

\newcommand{\verimap}{\textrm{VeriMAP}\xspace}

\title{{Enhancing Predicate Pairing with Abstraction for Relational Verification}\thanks{This 
work has been partially funded by INdAM-GNCS (Italy). 
\mbox{E.~De~Angelis,}
\mbox{F.~Fioravanti,} and A.~Pettorossi are research associates at IASI-CNR, Rome, Italy.}}

\author{Emanuele De Angelis\inst{1} \and Fabio Fioravanti\inst{1} \and\\ 
Alberto Pettorossi\inst{2,3} \and Maurizio Proietti\inst{3}}

\institute{DEC, University `G. d'Annunzio', Chieti-Pescara, Italy\\
\email{\{emanuele.deangelis,fabio.fioravanti\}@unich.it}\\
\and DICII, University of Rome `Tor Vergata', Italy\\
\email{pettorossi@info.uniroma2.it}\\
\and IASI-CNR, Rome, Italy \\
\email{maurizio.proietti@iasi.cnr.it}
}

\begin{document}
	
\maketitle

\pagestyle{headings}

\input{0_Abstract}

\section{Introduction}
\label{sec:Intro}
\input{1_Intro.tex}

\section{An Introductory Example}
\label{sec:IntroEx}
\input{2_IntroExample.tex}

\section{Constrained Horn Clauses over Numerical Domains}
\label{sec:Prelim}
\input{3_Prelim}

\section{Predicate Pairing with Abstraction}
\label{sec:PredicatePairing}
\input{4_PredicatePairing.tex}

\section{Experimental Evaluation}
\label{sec:Experiments}
\input{5_Experiments.tex}

\section{Related Work and Conclusions}
\label{sec:RelConcl}
\input{6_RWConclusions.tex}

\input{7_Bibliography.tex}

\end{document}

%% file: 0_Abstract.tex
\begin{abstract}
Relational verification is a technique that aims at proving properties 
that relate two different program fragments, or two different program runs. 
It has been shown that 
constrained Horn clauses (CHCs) can effectively be used for relational 
verification by applying a 
CHC transformation, called {\em predicate pairing}, 
which allows the CHC solver to infer relations among arguments 
of different predicates.
In this paper we study how the effects of the predicate pairing transformation 
can be enhanced by using various abstract domains based on linear arithmetic 
(i.e., the domain of convex polyhedra and some of its subdomains)
during the transformation.
After presenting an algorithm for predicate pairing with abstraction, we report on
the experiments we have performed on over a hundred relational verification problems by
using various abstract domains. The experiments have been performed by using 
the VeriMAP transformation and verification system, together with the 
Parma Polyhedra Library (PPL) and 
the Z3 solver for CHCs.
\end{abstract}

%% file: 1_Intro.tex
{\em Relational program properties} are properties 
that relate two different programs or two executions
of the same program.
Relational properties that have been studied in the literature
include 
program equivalence, 
non-interference for software
security, and relative correctness~\cite{Ba&11,Ben04,Lah&13}.

Recent work has advocated the use of {\em Constrained Horn Clauses} (CHCs)
for the verification of relational program properties~\cite{De&16c,Fe&14,MoF17}.
These methods translate a verification problem into a set of 
Horn clauses with constraints in a suitable domain (usually, Linear Arithmetic),
and then verify the satisfiability of those clauses by using SMT solvers
for Horn clauses, called here CHC solvers, such as Z3~\cite{DeB08} or Eldarica~\cite{Ho&12}.

The main difficulty encountered by CHC solvers when verifying relational
properties is that these solvers find 
models of {\it single} predicates expressed in terms of Linear Arithmetic constraints,
whereas the proof of relational properties often requires the discovery
of relations among arguments of {\it two $($or more$)$} distinct predicates.
To mitigate this difficulty, {\em Predicate Pairing} transforms a set of clauses
defining two predicates, say $p$ and $q$, into a new set of clauses defining
a new predicate, say $r$, equivalent to the conjunction of $p$ and $q$~\cite{De&16c}. 
Thus, when the CHC solver finds a model for the predicate~$r$, it discovers
relations among the arguments of $p$ and $q$.

In the approach presented in this paper we use Predicate Pairing together
with {\em Abstraction}.
{\rm Abstraction} is a technique that is often used in 
program analysis and transformation, and consists in mapping the 
concrete semantics of a program into an abstract domain, where some
program properties can more easily be verified~\cite{CoC77}. 
In the context of the verification of relational properties, 
Predicate Pairing {combined with} a basic form of abstraction 
has been introduced in a previous paper~\cite{De&16c}. 
In that paper, in fact, Predicate Pairing is performed 
by introducing new definitions whose bodies are made out of two 
atoms whose predicates are the ones to be paired, together with
the equalities between the arguments of these predicates, 
and these equality constraints
can be viewed as an abstraction 
into the domain of equalities.

Abstraction is also used by CHC {\it Specialization},  which is another transformation technique 
that has been proposed to increase the effectiveness of CHC solvers~\cite{De&14c,KaG17b}.
CHC Specialization propagates constraints through the clauses of the
program, and since propagation often causes strengthening of the constraints,
it is possible that by first specializing programs, 
the task of CHC solving is much facilitated. 
However, the impact of the specialization process very much depends on the choices
of the particular {\em abstract domain} and associated {\em widening operator}, 
which are used when specialized predicates are introduced or manipulated.

In this paper we address the problem of evaluating various combinations of
Predicate Pairing, Abstraction, and CHC Specialization for the
specific objective of verifying relational properties of programs.
In order to do so, we introduce a general algorithm for Predicate Pairing 
that is parametric with respect to the abstract constraint domain that is 
used. This domain is taken to be a subdomain of Linear Arithmetic, such as 
{\em Convex Polyhedra},
{\em Boxes}, {\em Bounded Differences}, and 
{\em Octagons}~\cite{Bag&08,Co&07,Min06}.
Our parametric Abstraction-based Predicate Pairing
algorithm generalizes the one that makes use of equalities between variables that has been used 
in a previous paper of ours~\cite{De&16c}.
We also consider a CHC Specialization algorithm that is parametric with
respect to the abstract constraint domain. Then we consider various compositions
of Abstraction-based Predicate Pairing and CHC Specialization, by varying
the abstract constraint domain used.

These compositions of transformations are applied to sets of CHCs
encoding relational properties of imperative programs.
The lesson one may learn from the results of our 
experiments is that the transformations achieving
the best results make use of constraint domains that are expressive enough
to describe relations between variables (such as, Bounded Differences and Octagons),
without requiring the higher precision of Convex Polyhedra.
Moreover, Abstraction-based Predicate Pairing essentially incorporates
the effect of CHC Specialization, and thus extra 
{specializations} 
(before or after Abstraction-based Predicate Pairing) are not 
{cost-effective}. 

The paper is organized as follows. In Section~\ref{sec:IntroEx} we present
an introductory example showing the usefulness of abstraction. 
Then, in Section~\ref{sec:Prelim} we present
the various abstract constraint domains, that is,
Convex Polyhedra, Boxes, Bounded Differences, and 
Octagons, and the operations defined on them.
In Section~\ref{sec:PredicatePairing} we present the Abstraction-based transformation 
techniques for CHCs and we prove that they preserve satisfiability (and unsatisfiability).
The implementation of the verification method, based on the Parma Polyhedra Library
for constraint manipulation~\cite{Bag&08} 
and on the Z3 solver for CHC satisfiability, 
and its experimental evaluation are 
reported in Section~\ref{sec:Experiments}. 
Finally, in Section~\ref{sec:RelConcl}, we
discuss the related work on program transformation and verification.

%% file: 2_IntroExample.tex

In this section we present the proof of equivalence  
of two imperative programs acting on {integer variables}. This equivalence 
can be proved by using our method based on the Predicate Pairing
strategy using a suitable abstract constraint domain. By using 
Predicate Pairing without abstraction the proof cannot be performed.

Let us consider the programs~$P1$ and~$P2$ shown in Figure~\ref{tab:looppipelining}, 
where program~$P2$ is obtained from program~$P1$ by applying 
a compiler optimization technique, called \textit{software pipelining}. 
This technique has the effect of allowing more parallelism during execution.

%

Software pipelining takes as input a program with a loop
and produces an improved program with a new loop 
whose instructions are taken from different iterations of the original loop.
Combined with other optimizations, 
software pipelining can produce loops whose instructions
have {no read/write} dependencies
and thus can be executed in parallel.
For example, in program~$P2$ 
derived by pipelining from program $P1$,
the dependency on {\tt x} 
in the instructions of the loop in $P2$ 
can be removed by:
(i)~introducing a fresh variable~{\tt u} initialized to the value of~{\tt x},
and 
(ii)~replacing~{\tt x} by~{\tt u} on the right-hand side of the assignments in the loop.
{After this replacement the resulting instructions 
`{\tt{y\,=\,y+u;}}' and `{\tt{a\,=\,a+1; x\,=\,u+a;}}' can be safely 
executed in parallel.}

\begin{figure}[ht]
\vspace{-10pt}
\begin{center}
\renewcommand{\baselinestretch}{.85} 
\begin{tabular}{c | c }

\hspace{8mm}\raisebox{14.5mm}{$P1:$} \begin{minipage}[h]{0.4\textwidth}
\vspace*{-2mm}\begin{verbatim}
   while (a < b) {
      x = x+a;  
      y = y+x;
      a = a+1;
   }
\end{verbatim}
\end{minipage}\hspace{-10mm}

& 

\hspace{4mm}\raisebox{14.5mm}{$P2:~~~$} \begin{minipage}[h]{0.5\textwidth}
\begin{verbatim}
if (a < b) {
   x = x+a;
   while (a < b-1) {
      y = y+x;
      a = a+1;
      x = x+a;  
   }
   y = y+x;
   a = a+1;
}
\end{verbatim}
\end{minipage}\hspace{-10mm}
\vspace*{-8pt}
\end{tabular}
\renewcommand{\baselinestretch}{1}  
\end{center}
\label{tab:looppipelining}
\caption{\rm {The input program $P1$ and the output program $P2$ obtained 
by applying software pipelining. 
}}
\end{figure}

The equivalence of programs $P1$ and $P2$ with respect to
the output value of~{\tt x},
can be expressed by the following clause~$F$: \nopagebreak

\smallskip
\noindent
$F\!:$~ $\textit{false} \leftarrow X1\!\neq\! X2,\  
   \textit{s\/}11(A,B,X,Y,X1,Y1),\ \textit{s\/}21(A,B,X,Y,X2,Y2) $

\smallskip
\noindent
where: (i)~predicates $\textit{s\/}11$ and $\textit{s\/}21$
represent the input/output relation of programs~$P1$ and~$P2$, respectively;
(ii)~$A$, $B$, $X$, $Y$
represent the values of the integer variables {\tt a}, {\tt b}, {\tt x}, {\tt y}
at the beginning of program execution; (iii)~$X1$, $Y1$ and $X2$, $Y2$
represent the values of the integer variables {\tt x} and  {\tt y} 
at the end of the execution of $P1$ and $P2$, respectively.
{In order to be processed by the constraint-based techniques presented in this paper,
which do not consider disequality ($\neq$), clause~$F$ is split into the following two clauses:

\smallskip
\noindent
$F1\!:$~ $\textit{false} \leftarrow X1\!\leq\! X2\!-\!1,\  
\textit{s\/}11(A,B,X,Y,X1,Y1),\ \textit{s\/}21(A,B,X,Y,X2,Y2) $

\noindent
$F2\!:$~ $\textit{false} \leftarrow X2\!\leq\! X1\!-\!1,\  
\textit{s\/}11(A,B,X,Y,X1,Y1),\ \textit{s\/}21(A,B,X,Y,X2,Y2) $}

\smallskip

The clauses defining $\textit{s\/}11$ and $\textit{s\/}21$ (and the predicates 
$\textit{s\/}12$, $\textit{s\/}22$, and $\textit{s\/}23$ on which they depend)
are reported below. (The unfamiliar reader may refer, for instance, to~\cite{De&17b}, 
where the technique for
generating clauses from imperative programs is presented.)
Note, for example, that predicates $\textit{s\/}12$ and $\textit{s\/}23$ 
correspond to the loops in programs~$P1$ and~$P2$, respectively.
{Note also that strict inequalities occurring in programs
(for instance, {\tt a\,<\,b} in program~$P1$) are
represented by using non-strict inequalities in clauses
($A\leq B\!-\!1$ in the first clause for 
$\textit{s\/}12$).}

\medskip
\noindent
$\textit{s\/}11(A,B,X,Y,X2,Y2) \leftarrow \textit{s\/}12(A,B,X,Y,A2,B2,X2,Y2). $

\noindent
$\textit{s\/}12(A,B,X,Y,A2,B2,X2,Y2) \leftarrow A\leq B\!-\!1, X1\!=\!A\!+\!X, Y1\!=\!Y\!+\!X1, $

\hspace{30mm} $A1\!=\!A\!+\!1,      \textit{s\/}12(A1,B,X1,Y1,A2,B2,X2,Y2).$

\noindent
$\textit{s\/}12(A,B,X,Y,A,B,X,Y) \leftarrow A\geq B.$

\medskip
\noindent 
$\textit{s\/}21(A,B,X,Y,X2,Y2) \leftarrow \textit{s\/}22(A,B,X,Y,A2,B2,X2,Y2).$

\noindent 
$\textit{s\/}22(A,B,X,Y,A2,B2,X2,Y2) \leftarrow A\leq B\!-\!1,\ X1\!=\!X\!+\!A,  $

\hspace{30mm} $\textit{s\/}23(A,B,X1,Y,A2,B2,X2,Y2).$

\noindent 
$\textit{s\/}22(A,B,X,Y,A,B,X,Y) \leftarrow A\geq B.$

\noindent 
$\textit{s\/}23(A,B,X,Y,A2,B,X,Y2) \leftarrow A\geq B\!-\!1,\ Y2\!=\!Y\!+\!X,\ A2\!=\!A\!+\!1.$

\noindent 
$\textit{s\/}23(A,B,X,Y,A2,B2,X2,Y2) \leftarrow A\leq B\!-\!2, \ Y1\!=\!Y\!+\!X,\ A1\!=\!A\!+\!1,   $

\hspace{30mm} $X1\!=\!X\!+\!A1, \     \textit{s\/}23(A1,B,X1,Y1,A2,B2,X2,Y2).$

\smallskip
\noindent
Let $P$ be the set of clauses defining $\textit{s\/}11$ and~$\textit{s\/}21$ and the
predicates on which they depend.
By proving the satisfiability of $\{F1, F2\} \cup P$,
we prove that programs $P1$ and~$P2$
produce identical values for  {\tt x}  as output,
when provided with the same input values.
{For reasons of simplicity we only consider the subset $\{F1\} \cup P$.
The satisfiability of the larger set can be proved by applying the same technique.}

Unfortunately, a state-of-the-art CHC solver such as Z3~\cite{DeB08}
fails to prove the satisfiability of $\{F1\} \cup P$.
This inability is due to the fact that Z3
computes models of {\em single} predicates expressed in terms of {\em linear} constraints,
and it can be shown that {\em non-linear} constraints among the arguments of
each predicate $\textit{s\/}11$ and~$\textit{s\/}21$ need be discovered 
to prove that the two atoms in the body of clause~$F1$ imply $X1\!>\! X2\!-\!1$. 


The Predicate Pairing strategy we have introduced in a previous paper~\cite{De&16c}
may help in overcoming this difficulty. 
By Predicate Pairing 
we may introduce new predicates defined in terms of two (or more) atoms,
and infer relations among arguments of distinct predicates.
Indeed, models of single predicates derived by Predicate Pairing, correspond
to models of pairs (that is, conjunctions) of predicates before Predicate Pairing.
Unfortunately, it is the case that also the set of clauses derived
from  $\{F1\} \cup P$
by `simple' Predicate Pairing, that is, without adding any constraint to the arguments
of the paired predicates,
cannot be proved satisfiable by Z3.

However, if we enhance our Predicate Pairing strategy by allowing in the new definitions
the addition of constraints taken from the abstract domain of Convex Polyhedra, 
then Z3 is indeed able to prove the  satisfiability
of the derived set of clauses, and hence of the equisatisfiable set $\{F1\} \cup P$.

To get this result, 
we have implemented in the \mbox{VeriMAP} transformation system \cite{De&14b}
a new version of the Predicate Pairing strategy
enhanced with abstraction techniques based on Convex Polyhedra
(as well as other abstract constraint domains such as Boxes, Bounded Differences, and Octagons).
In our example, Predicate Pairing enhanced with abstraction on Convex Polyhedra
introduces the following definitions
(variables are automatically renamed by \mbox{VeriMAP}):

\smallskip


\noindent
$\textit{pp\/}1(A,B,C,D,E,F,G,H,  A,B,C,D,I,J,K,L) \leftarrow G\leq K\!-\! 1, $

\hspace{10mm} $  \textit{s\/}12(A,B,C,D, E,F,G,H),\  \textit{s\/}22(A,B,C,D, I,J,K,L).          $


\noindent
$\textit{pp\/}2(A,B,C,D,E,F,G,H,  A,B,K,D,M,N,O,P) \leftarrow G\leq  O\!-\!1,\   A\leq  B\!-\!1,  $

\hspace{10mm} $  K=A\!+\!C,\  \textit{s\/}12(A,B,C,D, E,F,G,H),\   \textit{s\/}23(A,B,K,D, M,N,O,P).$

\smallskip

\noindent
and derives the following final set of clauses:

\smallskip

\noindent
{$\textit{false} \leftarrow X1\leq X2\!-\! 1,\  \textit{pp\/}1(A,B,C,D,E,F,X1,H, A,B,C,D,I,L,X2,N).$}

\noindent
$\textit{pp\/}1(A,B,C,D,E,F,G,H, A,B,C,D,I,J,K,L) \leftarrow G\leq K\!-\! 1, A\leq B\!-\! 1,  $

\hspace{10mm} $  M=A\!+\!C,\ \textit{pp\/}2(A,B,C,D,E,F,G,H, A,B,M,D,I,J,K,L).$

\noindent
$\textit{pp\/}2(A,B,C,D,E,F,G,H,A,B,K,D,M,N,O,P) \leftarrow  G\leq O\!-\! 1,\  A\leq B\!-\!2,   $
		  
\hspace{10mm}  $             K\!=\!A\!+\!C,\ R\!=\!A\!+\!1,\ T\!=\!A\!+\!C,\ S\!=\!D\!+\!T,\ X\!=\!A\!+\!1,\  W\!=\!K\!+\!X,$

\hspace{10mm}  $     Y=D\!+\!K, \         \textit{pp\/}2(R,B,T,S,E,F,G,H, X,B,W,Y,M,N,O,P).$

\smallskip
\noindent
{Now, the satisfiability of this set of clauses is trivial, and is easily checked by
Z3, because it contains no constrained facts 
(that is, clauses with only constraints in their body), and hence a model is obtained by
taking both \textit{pp\/}1 and \textit{pp\/}2 to be {\em false}.

The constraints occurring in the new predicate definitions 
are crucial for deriving a set of clauses without constrained facts.
Indeed, it can easily be shown that if we introduce
a predicate defined by a conjunction of predicates with distinct variables as 
arguments (that is, we perform `simple' Predicate Pairing), then 
by unfolding the definition of that predicate
we generate constrained facts derived from those of 
\textit{s\/}12 and \textit{s\/}22.

\smallskip
The equivalence of the two programs $P1$ and $P2$ with respect to
 the output value of~{\tt y} can be proved in a similar way.}

%
%
%
%


%% file: 3_Prelim.tex

Let us recall some basic notions about: (i)~abstract domains 
often used in static program analysis~\cite{CoC77}, and 
(ii)~constrained Horn clauses (CHCs).


We consider the abstract domain of {\em Convex $($Closed\/$)$ Polyhedra}~\cite{Bag&08,CoH78},
CP for short, whose {\em atomic constraints} are of the form 
$a_1 \cdot x_1\!+\ldots+a_n\cdot x_n  \leq  a,$
where  $a$'s  and $x$'s are real coefficients and variables, respectively.
A {\em constraint} $c$ is either \textit{true}, or \textit{false}, or an atomic constraint,
or a conjunction of constraints. %

%
%

Given a formula $F$, by $\forall (F)$ and $\exists (F)$
we denote its universal and existential closure, respectively.
By $\textit{vars}(F)$  we denote the set of variables occurring in $F$.
A constraint $c$ is  said to be {\em satisfiable} if $\textrm{CP} \models \exists (c)$.
Given two constraints $c$ and~$d$, we say that $c$ {\em entails} $d$, and we write $c \sqsubseteq d$, if 
$\textrm{CP}\models \forall (c\rightarrow d)$. We say that $c$ and $d$ are {\em equivalent} if 
$c \sqsubseteq d$ and $d \sqsubseteq c$.

We  also consider some other abstract constraint domains,
namely \textit{Octagons},  \textit{Bounded Differences}, and \textit{Boxes},
which are subdomains of \textit{Convex Polyhedra}
in the sense that they are defined 
by putting restrictions on the form of the polyhedra 
associated with the atomic constraints.
These abstract domains have all 
\textit{true} and \textit{false} as constraints 
and are closed under conjunction.

The atomic constraints of the {Octagons} domain
are 
inequalities of the form \mbox{$ a_1\cdot x_1  + a_2\cdot x_2 \leq  a$},
where $ a\!\in \!\mathbb{R}$ and  $a_i\! \in\! \{-1,0,1\}$, for $i=1,2$.
The atomic constraints of the {Bounded Differences} domain
are inequalities
of the form \mbox{$a_1\cdot x_1  + a_2\cdot x_2 \leq  a$},
where $ a\!\in\! \mathbb{R}$,  $a_i\! \in\! \{-1,0,1\}$, for $i=1,2$, and 
$a_1$ is different from $a_2$.
The atomic constraints of the {Boxes}  domain
are inequalities
of the form $ x\leq a$,
where $ a\! \in \!\mathbb{R}$.

Each abstract constraint domain~$D\!\subseteq\! {\rm{CP}}$ is endowed with some operators
that we now define. 
Let $c$ and $d$ be two constraints in~$D$, or $D$-constraints. 

The {\em least upper bound} 
operator
is a function   $\sqcup: D\!\times\! D\rightarrow D$
such that
(i)~$c \sqsubseteq c \sqcup d$, 
(ii)~$d \sqsubseteq c \sqcup d$ and
(iii)~for all $D$-constraints $e$, if $c \sqsubseteq e$ and $d \sqsubseteq e$, 
then $c \sqcup d \sqsubseteq e$.

A {\em widening operator} 
is a function   $\nabla: D\!\times\! D\rightarrow D$
such that   
(i)~$c \sqsubseteq c \nabla d$, 
(ii)~$d \sqsubseteq c \nabla d$, and
(iii)~for all chains $y_0 \sqsubseteq y_1 \sqsubseteq \ldots $,
the chain   $x_0 \sqsubseteq x_1 \sqsubseteq \ldots ,$
where $x_0 = y_0$ and, for $i\!>\!0$,  $x_{i+1} = x_i \nabla y_{i+1}$, is finite.

The {\em abstraction operator} 
for a subdomain $D$ of CP, 
is a function   $\alpha: \textrm{CP}\rightarrow D$
such that 
(i)~$c \sqsubseteq \alpha(c)$, and
(ii)~for all $D$-constraints $e$, if $c \sqsubseteq e$, then $\alpha(c) \sqsubseteq e$.

The {\textit{projection}} of a $D$-constraint \textit{c} onto a set \textit{X} of variables, 
denoted $c \Downarrow X$,
is a $D$-constraint $c'$, with variables in \textit{X},
which is equivalent to 
$\exists \textit{Y}. \textit{c}$, where $ \textit{Y} \!=\!  \textit{vars}(c) \! - \! X$.
Clearly, $c \sqsubseteq c'$.



An {\it atom} is a formula of the form $p(X_{1},\ldots,X_{m})$,
where $p$ is a predicate symbol different from `$\leq$' and 
$X_{1},\ldots,X_{m}$ are distinct variables.

A~{\it constrained Horn clause}  (or simply, a {\it clause}, or a CHC) is 
an implication of the form  
$A\leftarrow c, G$ (comma denotes conjunction), 
where the conclusion (or {\it head\/}) $A$ is either an atom or \textit{false}, 
the premise (or {\it body\/}) is the conjunction of
a constraint  $c$ and a (possibly empty) conjunction~$G$ of atoms. 
The empty conjunction is identified with~\textit{true}.

We assume that variables occur distinct in the atoms of a clause,
although in the examples, for ease of reading, 
we feel free to write clauses with repeated occurrences of the same variable.

A set $S$ of CHCs is said to be {\em satisfiable} if $S\cup \textrm{CP}$
has a model, or equivalently, $S\cup \textrm{CP}\not\models \textit{false}$
(as done above, we identify convex polyhedra with linear arithmetic formulas).

%
%

%% file: 4_PredicatePairing.tex
In this section we present an algorithm for transforming CHCs, 
called {\em Ab\-strac\-tion-based Predicate Pairing} (or {\it APP strategy}, for short),
which combines Predicate Pairing~\cite{De&16c} 
with abstraction operators acting on a given constraint domain (see Figure~\ref{fig:pairing}).
The APP transformation strategy preserves satisfiability of clauses and has
the objective of increasing the effectiveness of the satisfiability check
that is performed by the subsequent application of a CHC solver.

The APP transformation strategy
tuples together two or more predicates into
a single new predicate which is equivalent to their conjunction.
As shown in the example of Section~\ref{sec:IntroEx}, the
interaction between the Predicate Pairing technique
and the constraints among the variables
of the predicates paired together (or tupled together, 
if more than two), may ease the discovery of the relations 
existing among the arguments of the individual predicates.

The APP strategy is parametric with respect to: (i)~the 
abstract constraint domain which is considered, and 
(ii)~a {\it Partition} operator
that determines, given a clause, which are the atoms to be tupled together by
splitting any conjunction~$G$ of atoms in the body of the clause 
into $n\ (\geq\! 1)$ subconjunctions $G_1,\ldots,G_n$.
By choosing the abstract constraint domain and the Partition
operator in a suitable way, we can derive other
transformations, and in particular,
the  Predicate Pairing strategy introduced in a previous paper~\cite{De&16c},
{\em Linearization}~\cite{De&15c},
and CHC {Specialization}~\cite{De&14c,KaG17b}. For example, the
Predicate Pairing strategy is derived by taking the constraint domain to be the 
set of equalities between variables, and
the Partition operator to be the one which, given a clause, returns a suitable set of pairs 
of atoms in its body~\cite{De&16c}.

The APP strategy is realized by performing a
sequence of applications of the well-known {\em unfold/fold rules\/}~\cite{EtG96}.
In order to be self-contained, now we present the version of the Unfolding rule
used in this paper. The other rules will be presented when describing the APP
strategy. In the definition of the Unfolding rule and in the definitions of the 
other rules as well, we assume, without loss of generality, that: (i)~every
atom in a clause has distinct variables as arguments, and (ii)~two atoms have
pairwise disjoint sets of variables, and hence the relations among variables 
(including equalities) are explicitly written as constraints in the body of the clause.

\smallskip
\noindent
{\it Unfolding  Rule.} Let {\it P} be a set of clauses and $C$ be a clause 
of the form \linebreak$H\leftarrow c,L,A,R$, where $A$ is an atom 
and ${L}$ and ${R}$ are (possibly empty)
conjunctions of atoms. Let us consider the set
$\{A \leftarrow {c}_i,B_i \mid  i=1, \ldots, m\}$ 
made out of all the clauses in~$P$ whose head is $A$ (after renaming). 
By unfolding~$C$ w.r.t.~$A$ using~$P$, we derive the set of clauses
$\{({H}\leftarrow c,c_i,L,B_i,R) \mid i=1,\ldots, m\}$.

\smallskip

The APP strategy constructs a tree \textit{Defs} of {\em definitions}, that is, clauses
whose head predicates do not occur in the input set $P$ of clauses.
A definition $D$ is said to be a {\em child} of a definition $C$, and equivalently, $C$ is said to be 
the {\em parent} of~$D$, if~$D$ is introduced to fold a clause 
derived by unfolding from clause $C$.

The {\em ancestor} relation on \textit{Defs} is the 
{\em reflexive}, {\em transitive} closure of the parent relation.
\vspace{-3mm}

%
%
%
%
%
%

\begin{figure}[h!t]
	\noindent\hrulefill
	
	\noindent {\it Input\/}: A set $P \cup \{C\}$ of clauses
	where $C$ is a clause whose head predicate does not occur in $P$.
	
	\noindent  {\it Output\/}: A set $\textit{TransfCls}$ of clauses.
	
	\vspace*{-2mm}
	\noindent 
	\rule{30mm}{0.1mm}
	
	\vspace{-2mm}
	\begin{flushleft}\vspace{-2.mm}
		\noindent \textsc{Initialization}:~
		$\textit{InCls}:= \{\textit{C\/}\}$;
		~$\textit{Defs}$ is the tree made out of the root clause $C$ only;
		~$\textit{TransfCls}:= P$;
		
		\vspace*{.5mm}
		
		\noindent \textit{while}\, there is a clause~$C$ in \textit{InCls} of the form $H \leftarrow c, B$
		~\textit{do}

		\smallskip
		\hspace*{3mm}\begin{minipage}{118mm} 
			
			\hangindent=3mm
			\noindent $\bullet$ \textsc{Unfolding}: From clause $C$ derive a set $\textit{U}(C)$ of clauses by
			unfolding $C$ with respect to each atom occurring
			in its body using $P$;
			
			\smallskip
			
			\noindent $\bullet$ 
			    \textsc{Clause Deletion}: Remove from $U(C)$ all clauses with an unsatisfiable constraint;
			
			\smallskip
			\hangindent=3mm
			\noindent $\bullet$ {\textsc{Definition}\,\&\,\textsc{Folding}:}

			\hangindent=8mm
			\noindent
			\hspace*{3mm}\makebox[4mm][l]{\it for}\, every clause $E\in U(C)$ of the form
			$H \leftarrow d,\, G$ ~\textit{do}
			
			\hspace*{8mm}{\it Partition} the conjunction $G$ into $n\, (\geq\! 1)$ subconjunctions $G_1,\ldots,G_n$;
			
			\hspace*{8mm}{\it for} $i=1,\ldots,n$ {\it do}
			
			\hspace*{12mm}$d_i := \alpha(d)\Downarrow V_i$, where $V_i$ is the set of variables in $G_i$;
			
			\hangindent=11.5mm
			\noindent
			\hspace*{12mm}{\it if\/} in \textit{Defs} there is no
			clause $\textit{newp}_i(V_i) \leftarrow e_i, G_i$  
			such that $d_i\sqsubseteq e_i$ {\it then}
			
			\hspace*{16mm}{\it if\/} in \textit{Defs} there is an ancestor clause of $C$
			of the form  $\textit{newq}(V_i) \leftarrow f_i, G_i$
			
			\hspace*{16mm}{\it then} $D_i := (\textit{newp}_i(V_i)  \leftarrow a_i, G_i)$,
			where $a_i = f_i \nabla (f_i \sqcup d_i)$ 
			
			\hspace*{16mm}{\it else} $D_i := (\textit{newp}_i(V_i)  \leftarrow d_i, G_i)$;

			\noindent
			\hspace*{16mm}$\textit{InCls} := \textit{InCls}\cup\{D_i\}$; 
			\ \
			add $D_i$ as a child of $C$ in \textit{Defs};
			
			\hspace*{8mm}{\it end-for\/};
			
			\noindent
			\hspace*{8mm}$\textit{TransfCls}:=\textit{TransfCls}\cup  \{H \leftarrow d,\textit{newp}_1(V_1),\ldots,\textit{newp}_n(V_n)\}$;\\[-4mm]

			\hspace*{8mm}$U(C) := U(C) - \{E\}$; 
	
			\noindent
			\hspace*{3mm}{\it end-for\/};
			
			\vspace{1mm}
			\noindent $\textit{InCls}:=\textit{InCls}-\{C\}$;
			
		\end{minipage} 
		
		\vspace{1mm}
		
		\noindent \textit{end-while}
		
	\end{flushleft}
	
	\vspace*{-6mm}
	\noindent\hrulefill
	\vspace*{-2mm}
	\caption {{\em The Predicate Pairing with Abstraction algorithm $($\!APP strategy\/$)$}}
	\label{fig:pairing}\label{alg:pairing1}
	\vspace*{-5mm}
\end{figure}

Note that, by construction, 
every constraint, either $a_i$ or $d_i$, occurring in a new definition $D_i$
introduced during the \textsc{Definition}\,\&\,\textsc{Folding} phase,
belongs to the abstract constraint domain.

The APP strategy terminates if there exists an integer $k$ such that,
for any application of the Partition operator producing $G_1,\ldots,G_n$,
the size of each $G_i$ is bounded by $k$.
For instance, the Predicate Pairing strategy presented in a previous paper~\cite{De&16c},
guarantees that $k\!=\!2$ is such an integer bound.
Indeed, if this bound exists, by the properties of the widening operator,
the set of definitions that can be introduced by APP
(and hence the number of executions of the {\it while} loop of the
strategy) is finite.

Moreover, from well-known correctness results for the unfold/fold rules~\cite{EtG96},
it follows that the APP strategy preserves satisfiability.

\begin{theorem}[Soundness of Predicate Pairing with Abstraction]
	\label{thm:soundness} 
	Let the set $P\cup \{C\}$ of clauses be the input of the
	APP strategy. If the strategy terminates and returns a set {\it TransfCls}
	of clauses, then $P \cup \{C\}$ is satisfiable iff {\it TransfCls} is
	satisfiable.
\end{theorem}

A CHC Specialization strategy with Abstraction, which we call
{\em ASp strategy}, 
can be derived by instantiating the APP strategy. 
The ASp strategy is obtained by using the Partition operator that, given
a conjunction of atoms 
$A_1,\ldots,A_n$ in the body of a clause, 
returns $n$ subconjunctions, each 
consisting of a single atom~$A_{i}$, for $i\!=\!1,\ldots,n$
(that is, no predicate pairing is made).
It has been shown that CHC Specialization 
can be very useful for increasing the effectiveness of 
satisfiability checkers~\cite{De&14c,De&17b,KaG17b}.


In the next section we will present the results of the experiments
we have performed by combining the APP strategy and the ASp strategy in various ways. 
In particular, these results illustrate the role of abstract constraint domains 
for Predicate Pairing, and they also show 
the usefulness of the APP strategy for making
CHC solvers more effective when performing satisfiability checking.


%% file: 5_Experiments.tex
\newcommand{\un}{Universe\xspace}
\newcommand{\bx}{Boxes\xspace}
\newcommand{\bd}{BDS\xspace}
\newcommand{\oc}{OS\xspace}
\newcommand{\po}{\textrm{CP}\xspace}
\newcommand{\stdw}{\textrm{-H}\xspace}
\newcommand{\bagw}{\textrm{-B}\xspace}
\newcommand{\domainx}{\textit{Domain}\xspace x\xspace}
\newcommand{\inprogs}{\textit{InProbls}\xspace}
\newcommand{\outprogs}{\textit{OutProbls}\xspace}
\newcommand{\atime}{\textit{Avg}\textit{Time}$1$}
\newcommand{\inclsm}{\textit{InCls}\xspace}
\newcommand{\outcls}{\textit{OutCls}}
\newcommand{\sratio}{\textit{SizeRatio}}
\newcommand{\mysp}{\hspace*{2pt}}
\newcommand{\myspd}{\hspace*{1pt}}
\newcommand{\stime}{\textit{Avg}\textit{Time}$2$}
\newcommand{\sprogs}{\textit{SolvedProbls}}
In this section we present the results of the experimental evaluation we have 
performed for assessing the effectiveness and the efficiency of the APP strategy.

\begin{figure}[ht]
\vspace{-2mm}
\centering
\includegraphics[scale=1]{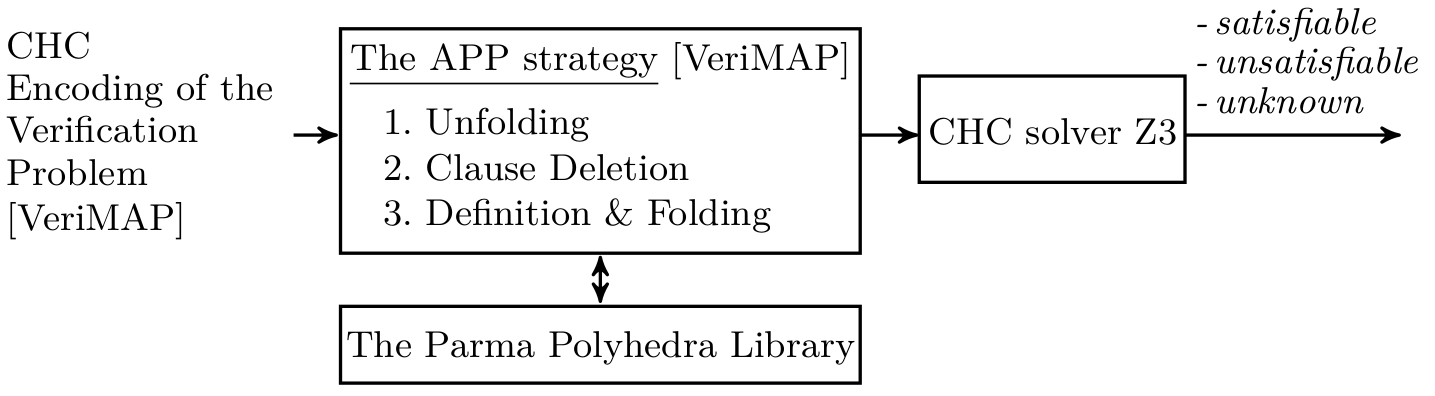}
\vspace{-5mm}
\caption{The verification process based on: 
(i)~the APP strategy, 
(ii)~the Parma Polyhedra Library, and 
(iii)~the CHC solver Z3.}
\label{fig:verproc}
\end{figure}

We have implemented that strategy using the~\verimap transformation 
system~\cite{De&14b} and the Parma Polyhedra Library~\cite{Bag&08}.
Then, in order to check the satisfiability of the clauses generated 
by the APP strategy we have used the~\textsc{Z3} solver~\cite{DeB08}.
The verification process is depicted in Figure~\ref{fig:verproc}.

\smallskip
\noindent
\textit{Implementation of the APP strategy.}
We have ported the \mbox{\verimap} tool {from SICStus Prolog 3.12.5} 
to SWI-Prolog 7.4.2 and we have extended its transformation engine so 
to use the abstract domains and the associated operations provided 
by the Parma Polyhedra Library (PPL).

In particular, we have considered the following abstract domains: 
(i)~{\em Universe} (that is, the whole $n$-dimensional space),
(ii)~{\em Boxes},
(iii)~{\em Bounded Differences} 
(also called Bounded Difference Shapes, or BDS, for short), 
(iv)~{\em Octagons} (also called Octagonal Shapes, or OS, for short), and 
(v)~{\em Convex Polyhedra}, 
together with the operations of projection, least upper bound, 
inclusion check {(that is, entailment)}, emptiness check 
(that is, satisfiability), and widening.
We have used Convex Polyhedra with the widening operator of
Halbwachs~\cite{Ha&97}~(CP\stdw), and with the widening operator of 
Bagnara et al.~\cite{Bag&05}~(CP\bagw).
Predicate Pairing with the Universe domain is the `simple' Predicate Pairing we 
mentioned in Section~\ref{sec:IntroEx}, that introduces new definitions having
no constraints in their body.

Since \verimap natively represents constraints using the Constraint Logic 
Programming (CLP) syntax, we had to use the facilities provided by PPL for 
constructing objects of the abstract domains starting from constraints 
represented in CLP, and vice versa, for constructing constraints 
represented in CLP starting from PPL objects.
We are currently working for overcoming the inefficiency due to these 
translations of representation.


\smallskip
\noindent
\textit{Benchmark suite.}
We have considered a benchmark suite consisting of 136 sets of CHCs, for a total
number of 1655 clauses, representing verification problems of various relational
properties, such as loop optimizations, equivalence, monotonicity, injectivity, 
functional dependency, and non-interference~\cite{Ba&11,Ben04,De&15c,De&16c,Fe&14}.

\smallskip
\noindent
\textit{Experiments.}
We have performed the following six experiments, where the parameter x stands 
for an abstract constraint domain, which is either Universe, or Boxes, or BDS, 
or OS, or CP\stdw, or CP\bagw:


\smallskip
\noindent
\begin{tabular}{lllll}
0.~ \textrm{Z3} \\ 

1.~ \textrm{ASp(x)\,;\,Z3}\\
2.~ \textrm{ASp(x)\,;\,APP(x)\,;\,Z3} \\
3.~ \textrm{ASp(x)\,;\,APP(x)\,;\,ASp(x)\,;\,Z3}\\

4.~ \textrm{APP(x)\,;\,Z3}\\

\multicolumn{2}{l}{5.~ \textrm{APP(Universe)\,;\,ASp(x)\,;\,Z3}}&
\end{tabular}
\smallskip

\noindent
Experiment 0 consists in running Z3 directly on the 1655 clauses that encode the 136 problems.
Z3 solved 28 problems, either positively or negatively, by providing the answer
`satisfiable' or `unsatisfiable', respectively, in an average time of~2.36 seconds 
per problem (see Frame~0 in Table~\ref{tab:eval}).

The other experiments with domain x have been performed as we now explain for 
Experiments~1 and~2 with domain
$\mathrm{x\!=\!OS}$ (see Frames~1 and~2, respectively, in Table~\ref{tab:eval}). 
By using \verimap we have run ASp(OS) on the initial 1655 clauses. 
Then, on the resulting 3540 clauses (with an average time of 0.73 seconds per problem),
{\em either} (Experiment~1) we have run Z3 that solved 28 problems 
(with an average time of 4.10 seconds per problem)
{\em or} (Experiment~2) by using again \verimap, we have run APP(OS) that produced 20361 clauses 
(with an average time of 4.60 seconds per problem). 
Finally, acting on these 20361 clauses, Z3 solved 121 problems out of 136
(with an average time of 3.90 seconds per solved problem).
In Column~{\it{SizeRatio\/}} we reported the increase in the number of clauses
due to the transformations performed by \verimap. 
In particular, ASp(OS) enlarged the size of about $2.14\, (=3540/1655)$ times, and
APP(OS) of about $5.83$ times.

During an experiment it may be the case that \verimap does not complete the execution of
the ASp(x) strategy, or the APP(x) strategy, within the given timeout of 300 seconds. In that case
the clauses of the problem for which the timeout occurred, are not passed to the
subsequent step of the experiment (and they are not considered in the computation of 
the average times and the size ratios).

We have used the Z3 4.5.0 solver with the Duality fixed-point engine~\cite{McR13} 
on an Intel Xeon CPU E5-2640 2.00GHz processor with 64GB of memory under the 
GNU/Linux 64 bit operating system CentOS~7.


\begin{table}[!ht]
	\centering
	\begin{tabular}{|ll|r|r|r|r|r||r|r|}
		
		\multicolumn{9}{l}{}\\[-8pt]\cline{1-9}
		\multicolumn{2}{|c}{} & \multicolumn{5}{c||}{\textbf{VeriMAP}} & \multicolumn{2}{c|}{\textbf{Z3}}\\
		\cline{1-9}
		\multicolumn{2}{|c|}{{\it Domain} x} & \inprogs  & \outprogs  & \outcls & \atime & \sratio & \sprogs & \stime\\ 
		\hline 
		%
		\multicolumn{9}{|l|}{\bf 0.\hspace{11mm} {\bf{\textbf{Z3}}}} \\\cline{3-9}
		& \hspace{-1mm}  & 136 & --- ~ &  --- ~ & --- ~ & --- ~ & 28\myspd &  2.36\myspd\\
		\hline \multicolumn{9}{l}{}\\[-10.4pt]\hline 
		
		
		\multicolumn{9}{|l|}{\bf 1.\hspace{11mm} {\bf{\textbf{ASp(x)\,;\,Z3}}}} \\\cline{3-9}
		& \bx           & 136 &  136 &  3111 & 0.67 & 1.88 & 29\myspd &  3.15\myspd\\
		& \bd           & 136 &  136 &  2629 & 0.66 & 1.59 & 28\myspd &  3.79\myspd\\
		& \oc           & 136 &  136 &  3540 & 0.73 & 2.14 & 28\myspd &  4.10\myspd\\ 
		& $\po{\stdw}$ & 136 &  136 &  3021 & 0.66 & 1.83 & 34\myspd &  3.95\myspd\\
		& $\po{\bagw}$ & 136 &  136 &  3633 & 0.69 & 2.20 & 36\myspd & 10.14\myspd\\ 
		\hline \multicolumn{9}{l}{}\\[-10.4pt]\hline 
		\multicolumn{9}{|l|}{{\bf 2.}\hspace{11mm} \textbf{Assuming ASp(x) already performed: ~ ~ ~ ~ ~ ~ ~~APP(x)\,;\,Z3}}  \\\cline{3-9}
		& \bx           & 136 &  134 & 27753 & 1.85 & 9.37 &  73\myspd & 2.20\myspd\\
		& \bd           & 136 &  136 & 12793 & 2.60 & 4.87 & 119\myspd & 3.69\myspd\\ 
		& \oc           & 136 &  134 & 20361 & 4.60 & 5.83 & 121\myspd & 3.90\myspd\\ 
		& $\po{\stdw}$  & 136 &  135 & 16193 & 3.09 & 5.39 & 113\myspd & 0.93\myspd\\
		& $\po{\bagw}$  & 136 &  127 & 12554 & 2.82 & 3.80 & 114\myspd & 3.65\myspd\\ 
		\hline \multicolumn{9}{l}{}\\[-10.4pt]\hline 
		\multicolumn{9}{|l|}{\bf 3.\hspace{11mm} \textbf{Assuming\ ASp(x)\,;\,APP(x)\  already performed: ~~ASp(x)\,;\,Z3}} \\\cline{3-9}
		& \bx           & 134 &  134 & 45970 & 2.57 & 1.66 &  77\myspd & 3.54\myspd\\
		& \bd           & 136 &  136 & 26683 & 3.30 & 2.09 & 121\myspd & 3.86\myspd\\ 
		& \oc           & 134 &  134 & 36871 & 4.89 & 1.81 & 119\myspd & 3.06\myspd\\ 
		& $\po{\stdw}$  & 135 &  135 & 31521 & 3.91 & 1.95 & 115\myspd & 2.05\myspd\\ 
		& $\po{\bagw}$  & 127 &  127 & 25495 & 4.59 & 2.03 & 112\myspd & 1.27\myspd\\
		\hline \multicolumn{9}{l}{}\\[-10.4pt]\hline 
		\multicolumn{9}{|l|}{\bf 4.\hspace{11mm} \textbf{APP(x)\,;\,Z3}} \\\cline{3-9}
		& \un~  & 136 &  136 &  4097 & 0.71 &  2.48 &  73\myspd & 2.00\myspd\\
		& \bx             & 136 &   136 & 20296 & 2.27 & 12.26 &  78\myspd & 2.01\myspd\\
		& \bd             & 136 &   136 &  8630 & 1.38 &  5.21 & 121\myspd & 2.45\myspd\\ 
		& \oc             & 136 &   135 & 13762 & 2.97 &  8.37 & 120\myspd & 1.77\myspd\\ 
		& $\po{\stdw}$   & 136 &   135 & 13823 & 2.59 &  8.40 & 110\myspd & 1.57\myspd\\ 
		& $\po{\bagw}$   & 136 &   131 & 11718 & 2.22 &  7.35 & 113\myspd & 2.19\myspd\\
		\hline
		\multicolumn{9}{l}{}\\[-10.4pt]\hline 
		\multicolumn{9}{|l|}{\bf 5.\hspace{11mm} \textbf{Assuming\ APP(Universe)\  already performed: ~~~ASp(x)\,;\,Z3}} \\
		\cline{3-9}
		& \bx           & 136 &   136 & 19932 & 2.23 & 4.87 &  74\myspd & 3.07\myspd\\
		& \bd           & 136 &   136 &  8387 & 1.46 & 2.05 & 120\myspd & 1.63\myspd\\ 
		& \oc           & 136 &   135 & 14065 & 2.93 & 3.46 & 118\myspd & 1.39\myspd\\ 
		& $\po{\stdw}$ & 136 &   135 & 14111 & 2.57 & 3.47 & 112\myspd & 1.44\myspd\\
		& $\po{\bagw}$ & 136 &   129 &  9831 & 2.41 & 2.53 & 113\myspd & 2.05\myspd\\
		\hline
	\end{tabular}
	\smallskip
	\caption{
		Frame~0 reports the results obtained by using Z3 on the input clauses.
		Frames~1--5 report the results obtained by transforming the clauses using 
		\verimap and then checking their satisfiability using Z3.
		Column~`\domainx' reports the chosen abstract domain.  
		Column~`\inprogs' (`\outprogs') reports the number of problems given as input to
		(produced as output by) \verimap.
		Columns~`\outcls' and `\atime' report the number of output clauses and 
		the average time taken by \verimap.
		Column~`\sratio' reports the value \outcls/\inclsm, where \inclsm is the number of 
		clauses of the problems for which \verimap terminates within the timeout. 
		Columns~`\sprogs' and `\stime' report the number of solved problems and the average time 
		taken by Z3. 
		The times are the CPU seconds spent in user mode.
		A timeout of 300 seconds has been set for each run of \verimap and Z3.
	}
	\label{tab:eval}
\end{table}

\smallskip
\noindent
\textit{Results.}
The results of the experiments are summarized in Table~\ref{tab:eval}. In order
to have a more detailed understanding of the results, for the Experiments~2, 3, and~5, 
we have adopted an incremental presentation in the sense that, for instance, 
Frame~2  presents the results for `APP(x) ; Z3', and the results concerning the first step `ASp(x)' 
are to be read in the first five columns of Frame~1 where the experiment `ASp(x) ; Z3' is presented.

First we  observe that 
the use of the ASp and APP strategies significantly increases
the number of problems which Z3 can solve (see Frames~0 and~2 of Table~\ref{tab:eval}). 
Indeed, for Experiment~2 with domain OS the number 
of problems solved by Z3 is 121, while Z3 alone solved 28 problems only.


Our results show that the use of abstract domain Boxes is not very effective.
Indeed, in Frame~2 we see that 73 problems only are solved (with respect to 121 for OS).
Similarly, in Frame~3, only 77 problems are solved (with respect to 121 for BDS). 
The poor performance of Boxes with respect to those of OS and BDS can be explained by the 
fact that Boxes (that is, intervals of individual variables) are not expressive enough to
represent relations among variables of programs. 
Hence, they are of little help for proving the given relational property.
Note, however, that if precision is further increased, from BDS and OS to Convex
Polyhedra, the performance does not increase. 
Indeed, the number of solved problems decreases because the computations required for
CP\stdw and CP\bagw are more expensive and the timeout limit is reached more often.

Looking at frames~1, 2, and~3 (Columns for Z3) we see that the main contribution to
the increase of efficacy in proving the desired properties is due to
the APP strategy (when used with a suitable domain, such as BDS, OS, CP\stdw, 
and CP\bagw), rather then the ASp strategy. 
Indeed, the comparison of Frames~2 and~3 (Column~\sprogs\xspace) 
shows that the use of ASp after APP does not {make any significant change in the
efficacy}.
Note also that applying ASp before APP for BDS and OS does not really pay off. 
Indeed, the comparison of Frame~2 with Frame~4 shows that more problems 
can be solved with lower verification time by directly applying the APP strategy 
starting from the clauses encoding the verification problems. 

The comparison between Frames~4 and~5 (Columns for~Z3) tells us that the effect of the 
APP(x) strategy, for a given abstract domain~x, can also be obtained in two steps: 
(i)~performing `simple' Predicate Pairing, that is, APP(Universe), and 
(ii)~applying ASp(x) which realizes the abstraction on the domain~x.

The large values in Column~\outcls\xspace show that in our implementation
more effort should be put 
in keeping the number of clauses generated by the 
APP and ASp strategies small, and indeed 
ongoing work is devoted to the design of new transformation techniques
that have that effect.

%% file: 6_RWConclusions.tex
We have proposed various ways of combining transformation and abstraction
techniques for constrained Horn clauses with the goal of verifying
relational properties of imperative programs.
To this aim we have presented Predicate Pairing and Specialization
algorithms that are parameterized with respect to a given
abstract constraint domain and its operators.
Then we have presented an extensive experimental evaluation of 
CHC satisfiability problems encoding relational verification problems.
Our experiments show that suitable combinations of transformations
and abstraction dramatically increase the effectiveness of the Z3 solver on the given
benchmark.
The most effective techniques combine Predicate Pairing and Abstraction
based on Bounded Differences or Octagons~\cite{Bag&05,Min06}, that is, constraint domains
that are quite simple, but expressive enough to capture the relations
between predicate arguments. 

Relational verification has been extensively studied, 
and still receives much attention as a relevant problem in the field of software 
engineering~\cite{Ba&11,Ch&12,Ci&14,De&16c,Fe&16,Fe&14,GoS08,Lah&13,LoM16,StV16,Ve&12,ZaP08}.
In particular, during the software development process it may be helpful to prove
that the semantics of a new program version has some specified relation 
with the semantics of an old version.

Among the various methods to prove relational properties, those by Mordvinov and
Fedyunkovich~\cite{MoF17} and by Felsing et al.~\cite{Fe&14}  are the most closely
related to ours. 
The method proposed in the former paper~\cite{MoF17} introduces the notion of CHC product 
(somewhat related to our Predicate Pairing strategy), that is,
a CHC transformation that synchronizes computations to improve the
effectiveness of the CHCs satisfiability check.
The latter method proposed by Felsing et al.~\cite{Fe&14} presents proof rules for relations
between imperative programs that are translated into constrained Horn clauses. 
The satisfiability of these clauses which entails the relation of interest, is then checked by 
state-of-the-art CHC solvers.

In the literature there are also partially automated approaches that 
reduce the problem of verifying relational
properties (in particular, program equivalence) to a standard verification 
problem by using some composition operator between imperative 
programs~\cite{Ba&11,Lah&13,StV16,ZaP08}.

The reuse of existing verification techniques for proving relational program properties can 
also be realized by first computing {\em summaries} of the two programs, that is,
relations between the input and output values of the programs,
and then proving relational properties of these summaries.
This approach is followed by: (i)~Ganty et al.~\cite{Ga&13}, 
whose method also supports  a restricted class of recursive programs, 
(ii)~Hawblitzel et al.~\cite{Ha&13},
who use automated theorem provers for proving  relative termination,
and (iii)~Lopes and Monteiro~\cite{LoM16}, 
who use (possibly nonlinear) integer polynomials
and recurrences for
summarizing loops of programs acting on integers.

A method that, like ours, is parametric with respect to the semantics of the 
programming languages in which the programs to be verified are written, is 
proposed by S.~Ciob\^ac\u{a} et al. \cite{Ci&14}, 
who present a language-independent deductive 
system for proving mutual equivalence of programs.

The Predicate Pairing technique we present in this paper is a descendant of well-known techniques for
logic program transformation, such as {\em Tupling}~\cite{PeP94} and {\em Conjunctive Partial
Deduction}~\cite{De&99}, which derive new predicates defined in terms of 
conjunctions of atoms. The goal of these techniques is the
derivation of efficient logic programs by: (i)~avoiding multiple traversals of data 
structures and repeated evaluations of predicate calls, and 
(ii)~producing specialized program versions that take into account partial information on the 
input values.
An integration of Conjunctive Partial Deduction and abstract interpretation,
called {\em Abstract} Conjunctive Partial Deduction, has also been presented
in the literature~\cite{Leu04}.
%
Recent work has shown that the extension of these transformation techniques 
to constrained Horn clauses can play a significant role in improving 
the effectiveness of CHC solvers for proving properties
of imperative programs, and in particular for verifying relational properties~\cite{De&15d,De&16c}.

The CHC Specialization strategy we consider in this
paper is a variant of specialization techniques
for (constraint) logic programs which have been proposed to support
program verification~\cite{Al&07,De&14c,De&15c,De&17b,Za&09,KaG17b,KaG17a,Me&07,Pe&98}.
However, these techniques are focused on the verification of partial or total
correctness of single programs, and not on the relational verification. 



%
%
%
%

%